\documentclass[twocolumn,preprintnumbers,prl,nobibnotes,nofootinbib]{revtex4}
\usepackage{graphicx}
\usepackage{amssymb,amsmath,amsfonts}
\usepackage[hypertex]{hyperref}

\def\lozenge{\boxit{\hbox to 1.5pt{\vrule height 1pt width 0pt \hfill}}}

\def\eg{{\it e.g.}}

\def\to{\rightarrow}

\newcommand{\dalam}{\raise-1mm\hbox{\large$\Box$}}

\newcommand{\beq}{\begin{equation}}
\newcommand{\eeq}{\end{equation}}

\begin{document}

\pagestyle{plain}

\preprint{ANL-HEP-PR-07-97, MIT-CTP-3906, SLAC-PUB-12990}
\

\vskip 1cm
\title{The LHC Inverse Problem, Supersymmetry, and the ILC}

\author{{C.F. Berger}\footnote{This work is
supported in part by funds provided by the U.S. Department of Energy
(D.O.E.) under cooperative research agreement DE-FC02-94ER40818.}}
\email{cfberger@mit.edu}
\affiliation{Center for Theoretical Physics,
Massachusetts Institute of Technology, Cambridge, MA 02139, USA}
\author{{J.S. Gainer}\footnote{Work supported in part
by the Department of Energy, Contract DE-AC02-76SF00515.}}
\email{jgainer@slac.stanford.edu}
\affiliation{Stanford Linear Accelerator Center, 2575 Sand Hill Rd.,
Menlo Park, CA  94025, USA}
\author{{J.L. Hewett}$^\dagger$}
\email{hewett@slac.stanford.edu}
\affiliation{Stanford Linear Accelerator Center, 2575 Sand Hill Rd.,
Menlo Park, CA  94025, USA}
\author{{B. Lillie}\footnote {Research
 supported in part by the US Department of Energy under contract
 DE--AC02--06CH11357}}
\email{lillieb@uchicago.edu}
\affiliation{High Energy Physics Division, Argonne National Laboratory,
Argonne, IL 60439, USA}
\affiliation{Enrico Fermi Institute, University of Chicago, 5640 South Ellis Avenue,
Chicago, IL 60637, USA}
\author{{T.G. Rizzo}$^\dagger$}
\email{rizzo@slac.stanford.edu}
\affiliation{Stanford Linear Accelerator Center, 2575 Sand Hill Rd.,
Menlo Park, CA  94025, USA}


\begin{abstract}
We address the question whether the ILC can resolve the LHC
Inverse Problem within the framework of the MSSM.
We examine 242 points in the MSSM parameter space which
were generated at random and were found to give indistinguishable
signatures at the LHC. After a realistic simulation including full
Standard Model backgrounds and a fast detector simulation, we
find that roughly only one third of these scenarios lead to
visible signatures of some kind with a significance $\geq 5$
at the ILC with $\sqrt s=500$ GeV.  Furthermore, we examine
these points in parameter space pairwise  and find that only
one third of the pairs are distinguishable at the ILC at $5\sigma$.
\end{abstract}
\maketitle


The exploration of the Terascale begins in earnest next year with the start
of operations at the LHC. Although the Standard Model (SM) does a respectable
job at describing present data, we know that it is incomplete for many reasons.
One can argue persuasively that to address these issues new physics must appear
at the Terascale. Once new physics is found, our next goal will be to determine
the nature and detailed structure of the underlying framework from which it arose.
Can this be done uniquely using data from the LHC?  The answer to this question
has been recently quantified by Arkani-Hamed, Kane, Thaler and Wang
(AKTW)~{\cite {Arkani-Hamed:2005px}}, who demonstrated what has come
to be known as the LHC Inverse Problem.  These authors studied a restricted and
very specific `simple' form of new physics given by the Minimal Supersymmetric
Standard Model (MSSM). By performing a scan of MSSM parameter space,
they showed that specific signatures observed in LHC experiments can
not be uniquely mapped back to distinct parameter space points,
hereafter referred to as models, within the MSSM. The purpose of the present
Letter is to outline an investigation of the capability of the International
Linear Collider (ILC) to resolve this issue of degeneracy within the MSSM
framework arising from the LHC Inverse Problem. Such an analysis, though of
interest in its own right, provides a unique opportunity to make a detailed
study of the signals and backgrounds for hundreds of Supersymmetric (SUSY) models.
From this we can re-examine our basic assumptions and prejudices about SUSY analyses at the ILC.

An outline of our procedure is as follows{\footnote{For full details of our analysis see Ref. {\cite{us}}.}}:
($i$) We obtained the set of model pairs from AKTW which led to identical signatures at the LHC. After some
filtering, this consists of 242 models in 162 degenerate pairs with some models appearing in several pairs. ($ii$) We
next generated signal (S) `data' via Monte Carlo techniques for each of these models employing both
PYTHIA {\cite {Sjostrand:2006za}} and CompHEP {\cite {Boos:2004kh}} and employing a realistic
beamspectrum {\cite {Guinea}} specific to the ILC design. We assumed an ILC with a center
of mass energy of $\sqrt s=500$ GeV and an integrated luminosity of $500~\mbox{fb}^{-1}$ which was split equally
between samples with $80\%$ left- and right-handed electron beam polarization. Positron polarization was not
included. ($iii$) We obtained two, statistically independent SM background sets (B) which were generated by
T.Barklow {\cite {TimB}} and employ full tree-level matrix elements using WHIZARD/O'Mega {\cite {WHIZARD}}. These included
all ($>1000$) processes of the form $2\to 2,\,2\to4,$ and $2\to 6$ final states that can arise from
$e^+e^-, \gamma e^\pm$ and $\gamma \gamma$ collisions. The use of full matrix elements here is important as
they generally produce larger backgrounds with longer tails in kinematic
distributions than does, \eg, PYTHIA. ($iv$) We piped
both signal and background through the java-based SiD {\cite {SiD}} fast detector simulation, org.lcsim {\cite {lcsim}},
to take realistic detector effects into account. The procedure outlined above corresponds to a realistic scenario
of a first scan of the MSSM signature space at the ILC.

At this stage, we first ask what are the numbers and types of SUSY particles
that are kinematically accessible at the ILC in each of the 242 AKTW models?
The answer is presented in Table~\ref{finalstates}; here
we see that many more sparticles are accessible at 1 TeV than at 500 GeV, which is a good argument for going
to higher energies as soon as possible.

\begin{table}
\centering
\begin{tabular}{|c|c|c|} \hline\hline
Final State & 500 GeV & 1 TeV  \\ \hline
$\tilde e_L^+ \tilde e_L^-$                     &  9  & 82   \\
$\tilde e_R^+ \tilde e_R^-$                     &  15 & 86   \\
$\tilde e_L^\pm \tilde e_R^\mp$                 &  2  & 61  \\
$\tilde \mu_L^+ \tilde \mu_L^-$                 &  9  & 82  \\
$\tilde \mu_R^+ \tilde \mu_R^-$                 &  15 & 86   \\
Any selectron or smuon                          &  22 & 137  \\
$\tilde \tau_1^+ \tilde \tau_1^-$               &  28 & 145   \\
$\tilde \tau_2^+ \tilde \tau_2^-$               &  1  & 23  \\
$\tilde \tau_1^\pm \tilde \tau_2^\mp$           &  4  & 61  \\
$\tilde \nu_{e\mu}\tilde \nu_{e\mu}^*$          &  11 & 83   \\
$\tilde \nu_\tau \tilde \nu_\tau^*   $          &  18 & 83   \\
$\tilde \chi_1^+ \tilde \chi_1^-$               &  53 & 92   \\
Any charged sparticle                           &  85 & 224  \\
$\tilde \chi_1^\pm \tilde \chi_2^\mp$           &  7  & 33  \\
$\tilde \chi_1^0 \tilde \chi_1^0$               &  180& 236    \\
$\tilde \chi_1^0 \tilde \chi_1^0$ only          &  91 &  0  \\
$\tilde \chi_1^0 + \tilde \nu$ only             &  5  &  0 \\
$\tilde \chi_1^0 \tilde \chi_2^0   $            &  46 &  178  \\
  Nothing                                       &  61 &  3    \\ \hline\hline
\end{tabular}
\caption{Number of models, out of 242, which have a given final
state kinematically accessible at $\sqrt s=500$ GeV and 1 TeV.
Note that for the 500 GeV machine 96/242 models have
only LSP or neutral pairs accessible while 61/242 models have no SUSY particles accessible.}
\label{finalstates}
\end{table}

The next issue to address is how many of the kinematically
accessible sparticles in a given model actually produce an
observable signal at the ILC. Certainly we cannot distinguish
models which are invisible! To this end, we performed a number of
detailed analyses to search for the selectrons, smuons, staus,
sneutrinos, lightest charginos and the lightest two neutralinos in
each of the AKTW models, the details of which are given elsewhere
{\cite {us}}. An important point here has to do with choosing the
kinematic cuts to enhance the signal over background;
many such sets of cuts have been developed in the literature
for ILC SUSY analyses {\cite {ILClit,Weiglein:2004hn}}. These are
usually based on a few representative benchmark models, such as SPS1a'
{\cite {{Allanach:2002nj,SPS}}}, which have very large SUSY production
cross sections
and many spartners kinematically accessible. Since the AKTW models arise from a
generalized parameter scan, these existing cuts are not always
adequate for observing our SUSY signals, \eg, in some cases the existing cuts
completely removed our signal, while in other cases they did not sufficiently
reduce the SM background. We thus developed a
set of cuts that work more or less uniformly throughout the
parameter space. This led in many cases to increased backgrounds
compared with those found for the specialized points studied in
the literature.

In order to ascertain the significance, $\cal S$, of the model signal
for any given search analysis we employ the
likelihood ratio method, based on a Poisson statistics likelihood
function, and use both of our statistically
independent background sets as input. As is traditional, to claim
that sparticle production is observable, we require that ${\cal S} >5$.

\begin{figure}[htbp]
\centerline{
\includegraphics[width=7.5cm,angle=0]{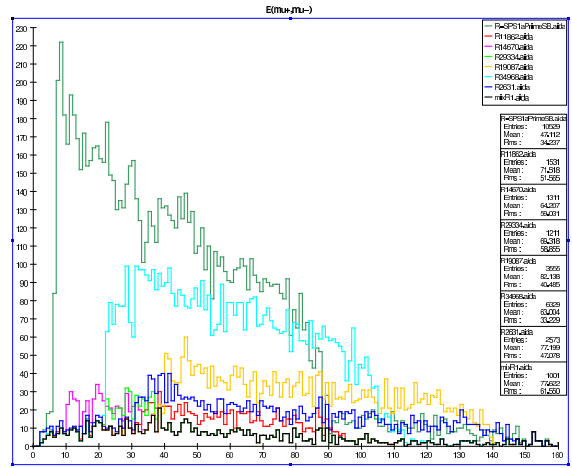}}
\vspace*{0.1cm}
\caption{Muon energy distribution: the number of events/2 GeV bin (combined signal and background) after
imposing selection  cuts for several representative models, with right-handed
electron beam polarization, assuming
an integrated luminosity of 250 fb$^{-1}$. For comparison, SPS1a' is shown as the histogram with the largest
number of events. The SM background is represented by the black histogram.}
\label{fig1}
\end{figure}

An example of one of our analyses is shown in Fig.~\ref{fig1}. Here we display
 the single muon energy distribution from the
production and decay of smuon pairs at the ILC; the specific process of relevance here is $e^+e^- \to
\tilde \mu^+\tilde \mu^- \to \mu^+\mu^-+2\tilde \chi_1^0$. Since
here $\chi_1^0$ is the lightest SUSY particle (LSP) and
is both stable and neutral, it appears in the detector as missing energy
and thus only the properties of the final state muons
can be measured experimentally.  Two things are apparent from this
figure: ($i$) Although the backgrounds are
relatively low due to judicious cuts, there is a wide range for the ratio S/B as
the MSSM parameter space is scanned.
Beam polarization plays an important role here.
($ii$) S/B for SPS1a' is significantly larger than in any of the
hundreds of models in our sample; this result is generally found to
hold in all of our other sparticle search analyses as
well.

\begin{figure}[htbp]
\centerline{
\includegraphics[width=5.7cm,angle=90]{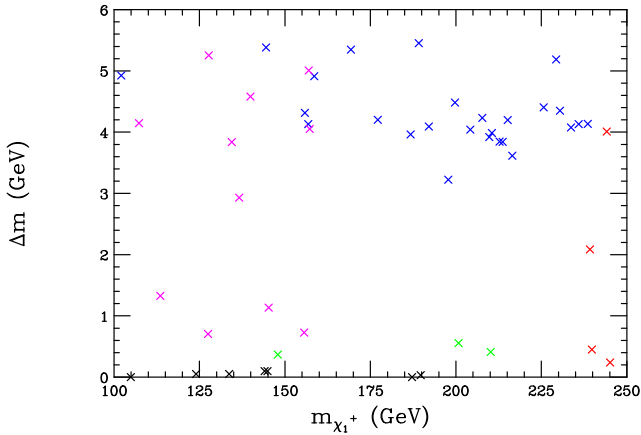}}
\vspace*{0.1cm}
\caption{Distribution of the chargino-LSP mass difference versus the chargino
mass for 51 of the AKTW models with $\Delta m_{\tilde \chi}<6$ GeV for the
$\tilde\chi_1^\pm$ states that are kinematically accessible at $\sqrt s=500$ GeV.
The blue crosses represent models that are observable in our suite of analysis channels
based on the $\tilde\chi_1^\pm$ decay via off-shell $W$ bosons.  The green
crosses correspond to models that are only visible in the radiative chargino
production analysis channel, while the magenta ones represent models
that yield observable signals in both the radiative and off-shell $W$
channels.  The black crosses are models that are visible in the stable
chargino analysis.  The red points are the 4 models where the
$\tilde\chi_1^\pm$ state is {\it not} observable in any of our
analysis channels, essentially due to phase space restrictions.}
\label{distribution}
\end{figure}

The situation for other spartners, such as the lightest chargino $\tilde \chi_1^+$,
is more complicated as
the observable signature strongly depends on the mass splitting with the lightest neutralino,
$\Delta m= m_{\tilde \chi_1^+}-m_{\tilde \chi_1^0}$. If $\Delta m$ is sufficiently large, one can observe the
decay of pairs of $\tilde \chi_1^+$'s through on- or off-shell $W$'s into jets and/or leptons. For smaller
values of
$\Delta m$, these decay products are soft, so we must
tag the final state by requiring the associated production of a high-$p_T$
photon {\cite {Gunion:2001fu,Riles:1989hd}}.
When $\Delta m < m_\pi$, $\tilde \chi_1^+$ is effectively stable, decaying outside the detector,
 and we
perform a search for long-lived charged
particles {\cite {{Martyn:2007mj}}}. By combining these various
techniques, we can cover the entire MSSM parameter space fairly well as
shown in Fig.~\ref{distribution}, where we find that only 4 out of the
53 models with accessible $\tilde\chi_1^\pm$ are not observable at the ILC.

Another potentially difficult case is stau pair production,  $e^+e^- \to \tilde \tau^+\tilde \tau^- \to
\tau^+\tau^-+2\tilde \chi_1^0$ since the final state $\tau$'s themselves decay and need to be identified in the
detector. In our analysis, we focus on the hadronic decays of taus into pions,
$\tau \rightarrow \pi \nu_\tau; \,
\tau \rightarrow \rho \nu_\tau \rightarrow \pi^{\pm} \pi^0 \nu_\tau; \,
\tau \rightarrow 3 \pi \nu_\tau$, the latter being a 3-prong jet, and also include the $e,\mu$ leptonic decays of the
$\tau$ as these make up a substantial branching fraction.
As an {\it alternative} possibility, we allow leptonic tau decays into muons, but reject
taus that decay into electrons.  This significantly reduces contamination from photon-induced backgrounds,
at the price of reducing the signal by roughly 30\%.
The reason for rejecting events with electrons is to reduce
contamination from gamma-induced Standard Model processes not involving taus.
The following main background, which has large missing energy,
and thus mimics stau production, is removed by the alternative tau ID
method: $e e \gamma \gamma \rightarrow e (e) \mu (\mu) + E^{\mbox{\tiny miss}}$,
where the not reconstructed final states are indicated by the brackets ( ),
and a beam electron or positron is falsely identified as a tau decay
product.
Better forward muon ID capabilities would remedy this situation because
the second muon would then be captured by the detector as well.

\begin{figure}[htbp]
\centerline{
\includegraphics[width=7.5cm,angle=0]{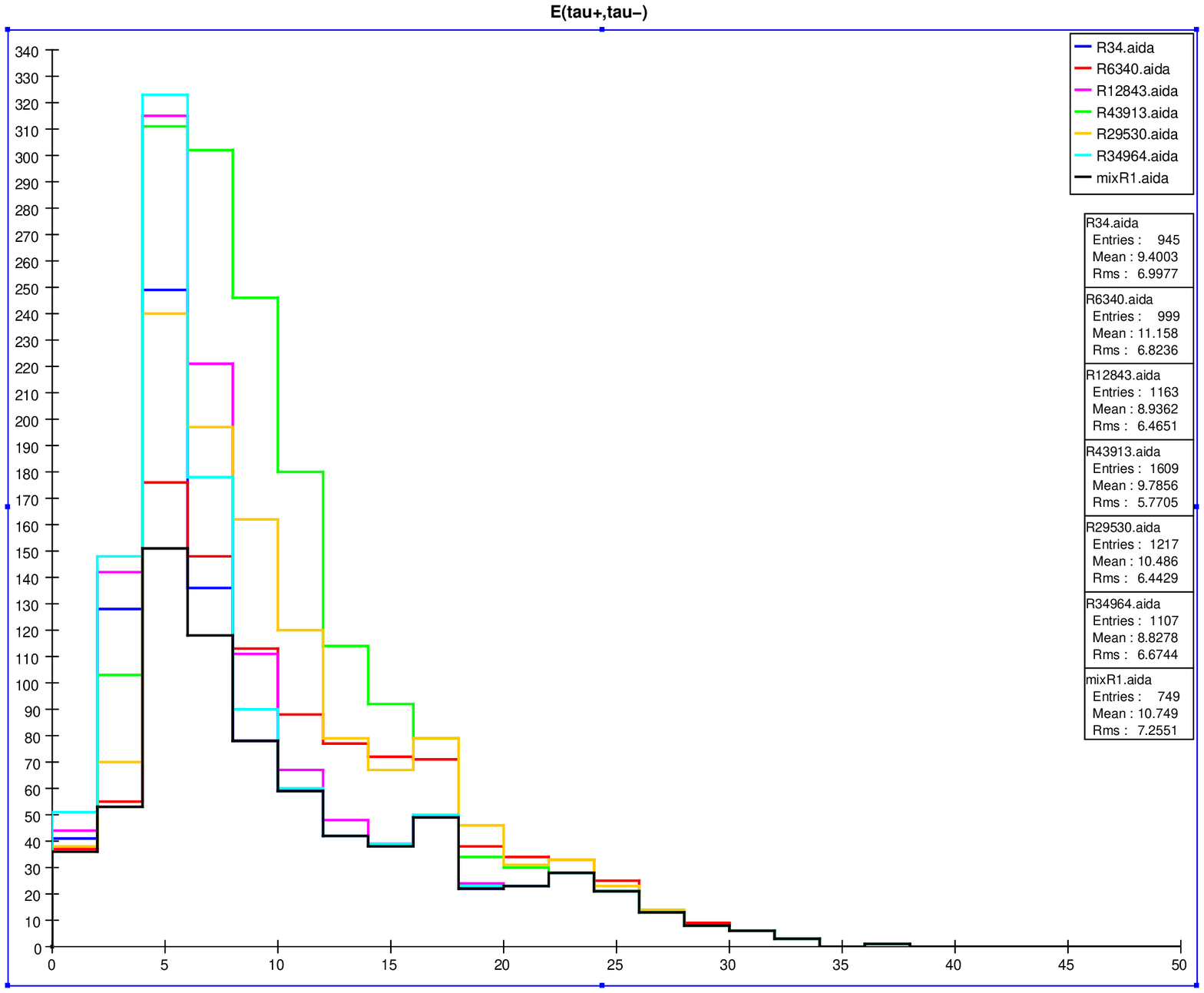}}
\centerline{
\includegraphics[width=7.5cm,angle=0]{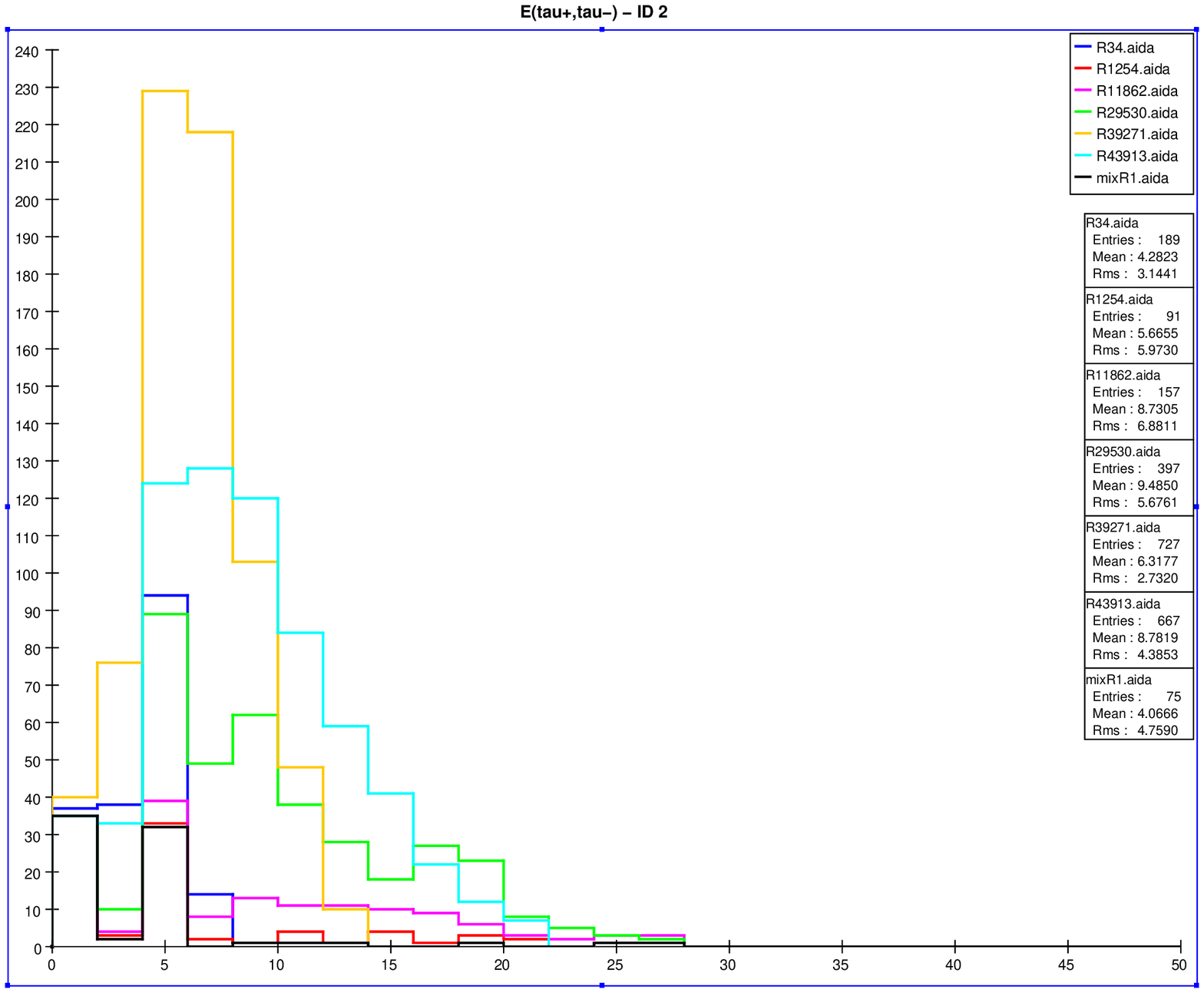}}
\vspace*{0.1cm}
\caption{Reconstructed $\tau$ energy distribution: the number of events/2 GeV bin (combined signal and background)
after imposing selection cuts for several representative models, with right-handed beam polarization, assuming
an integrated luminosity of 250 fb$^{-1}$. The bottom panel corresponds to removing
the electron final state in
$\tau$ decays in order to avoid beam backgrounds as discussed in the text.
The SM background is represented by the
black histograms.}
\label{figstaus}
\end{figure}

The results from both the standard and alternative $\tilde \tau$ analyses are shown in Fig.~\ref{figstaus}.
While stau production is certainly observable above background in both cases,
the alternative technique nearly completely eliminates the background.
Augmenting the detector with muon identification (ID) capabilities at
lower angles could reduce the $\gamma$-induced background without having to pay the
price of introducing a restricted tau identification. Apparently a significant portion of the AKTW models
have relatively low stau signal rates, and an improved muon ID
capability can be crucial if in fact this portion of the SUSY parameter space is realized in nature.
Note that while 21/28 models with kinematically accessible $\tilde \tau$'s are visible using the alternative
ID, the number is reduced to only 13/28 if the standard method, which includes electron
final states, is employed.

Table~\ref{visibleparticles} summarizes the results of our individual model searches.
Our analyses do well
at finding charged spartners but not so well with neutral ones. We find that 78/85 models with at least
 one charged
spartner kinematically accessible are visible, but only 17/96 models with
only neutral spartners kinematically accessible are observed. This translates into
82 visible models out of 161 models with kinematically accessible spartners, and
82 models that are visible out of the total 242 AKTW models.

\begin{table}
\centering
\begin{tabular}{|c|c|} \hline\hline
Particle & Number Visible   \\ \hline
$\tilde e_L$                       &  8/9     \\
$\tilde e_R$                       &  12/15    \\
$\tilde \mu_L$                     &  9/9     \\
$\tilde \mu_R$                     &  12/15     \\
$\tilde \tau_{1}$                  &  21/28  \\
$\tilde \nu_{e,\mu}$               &  0/11     \\
$\tilde \nu_\tau$                  &  0/18    \\
$\tilde \chi_1^\pm$                &  49/53     \\
$\tilde \chi_1^0$                  &  17/180     \\
$\tilde \chi_2^0$                  &  5/46     \\ \hline\hline
\end{tabular}
\caption{Number of models, at $\sqrt s=500$ GeV, which have a given final state particle visible above
the SM background with a significance ${\cal S}>5$ divided by the number of models with the same particle
kinematically accessible.}
\label{visibleparticles}
\end{table}

Finally, now that we know which models lead to observable signatures at the ILC, we can ask how well the ILC
performs at distinguishing pairs of models that gave degenerate signals at the LHC.
To this end we perform a $\chi^2$ comparison
of the signal plus background histograms for each of our analyses for each of the model pairs. Specifically, we
construct a $\chi^2$ of the form $\chi^2(S1+B1,S2+B2)$ where $S1(2)$ are the pure signal samples for the two
models being compared,
and $B1(2)$ are our two independent background samples. We find that we can distinguish, at the $5(3)\sigma$
confidence level, 57(63)/72 pairs of models where at least one of the models has a charged spartner
kinematically accessible, but we fail
completely when both models being compared have only neutral spartners accessible. Thus,
 out of all of the AKTW model pairs, our results show that
 57(63)/162 can be distinguished at the ILC $5(3)\sigma$.

From this analysis it is clear that the ILC with the SiD detector does a respectable job in observing
charged sparticles that are kinematically accessible, and in distinguishing models that contain
such particles. The major weakness, beyond the restricted kinematic reach, is in the neutral spartner sector.
This problem might be resolved by employing positron beam polarization {\cite {MoortgatPick:2005cw}} as
well as by implementing
more sophisticated analyses. In order to observe more sparticles in all models,
 an upgrade to $\sqrt s=1$ TeV as soon as possible is desirable.

\acknowledgments
We are indebted to Tim Barklow for extraordinary help with the Standard Model
background and for many invaluable discussions.


\begin{thebibliography}{99}
\bibitem{TimB}
The SM backgrounds we employ here have been generated by T.~Barklow.
\end{thebibliography}
\end{document}